\documentclass[journal, onecolumn, 11pt]{IEEEtran}

\usepackage{cite}

\ifCLASSINFOpdf
  \usepackage[pdftex]{graphicx}
  \graphicspath{{../figure/}}
  \DeclareGraphicsExtensions{.pdf,.jpeg,.png}
\else
  \usepackage[dvips]{graphicx}
  \graphicspath{{../figure/}}
  \DeclareGraphicsExtensions{.eps}
  
\fi

\usepackage[cmex10]{amsmath}
\usepackage[hidelinks,colorlinks=true,linkcolor=blue,citecolor=red, urlcolor=black]{hyperref}
\usepackage{booktabs}
\usepackage{verbatim}
\usepackage[dvipsnames]{xcolor}
\usepackage{dirtytalk} 
\usepackage{array}
\usepackage[shortcuts]{extdash}
\usepackage{tabularx}
\usepackage{fancyhdr}
\usepackage{wrapfig}
\usepackage{floatflt}

\usepackage{multirow}

\begin{document}
%
\title{The Smart Mask: Active Closed-Loop Protection against Airborne Pathogens}

\author{\IEEEauthorblockN{Naren Vikram Raj Masna, Rohan Reddy Kalavakonda, Reiner Dizon, Anamika Bhuniaroy, Soumyajit Mandal, and Swarup Bhunia\\ University of Florida, Gainesville, FL, USA\\Email: swarup$@$ece.ufl.edu}}

\maketitle

\begin{abstract}
Face masks provide effective, easy-to-use, and low-cost protection against airborne pathogens or infectious agents, including SARS-CoV-2. There is a wide variety of face masks available on the market for various applications, but they are all passive in nature, i.e., simply act as air filters for the nasal passage and/or mouth. In this paper, we present a new ``active mask'' paradigm, in which the wearable device is equipped with smart sensors and actuators to both detect the presence of airborne pathogens in real time and take appropriate action to mitigate the threat. The proposed approach is based on a closed-loop control system that senses airborne particles of different sizes close to the mask and then makes intelligent decisions to reduce their concentrations. This paper presents a specific implementation of this concept in which the on-board controller determines ambient air quality via a commercial particulate matter (PM) sensor, and if necessary activates a piezoelectric actuator that generates a mist spray to load these particles, thus causing them to fall to the ground. The proposed system communicates with the user via a smart phone application that provides various alerts, including notification of the need to recharge and/or decontaminate the mask prior to reuse. The application also enables a user to override the on-board control system and manually control the mist generator if necessary. Experimental results from a functional prototype demonstrate significant reduction in airborne PM counts near the mask when the active protection system is enabled. 
\end{abstract}

\IEEEpeerreviewmaketitle

\section{Introduction}
The world has been witnessing the increasing spread of COVID-19 since the beginning of 2020. This novel virus has brought everyone's lives to a standstill and the economy to its knees. Although most people are actively following social distancing norms, proper hygiene, and other preventive measures, it is likely that normal day-to-day life will continue to be affected until an effective vaccine is developed. To mitigate this situation and  return to some semblance of normalcy, we believe there is a need for preventive methods that actively combat the virus instead of providing passive protection (e.g., physical barriers). Implementing such improved methods requires smart devices that can detect, quantify, and actively eradicate viruses and other pathogens.
 
\begin{figure}[t]
\centering
\includegraphics[width=0.5\textwidth]{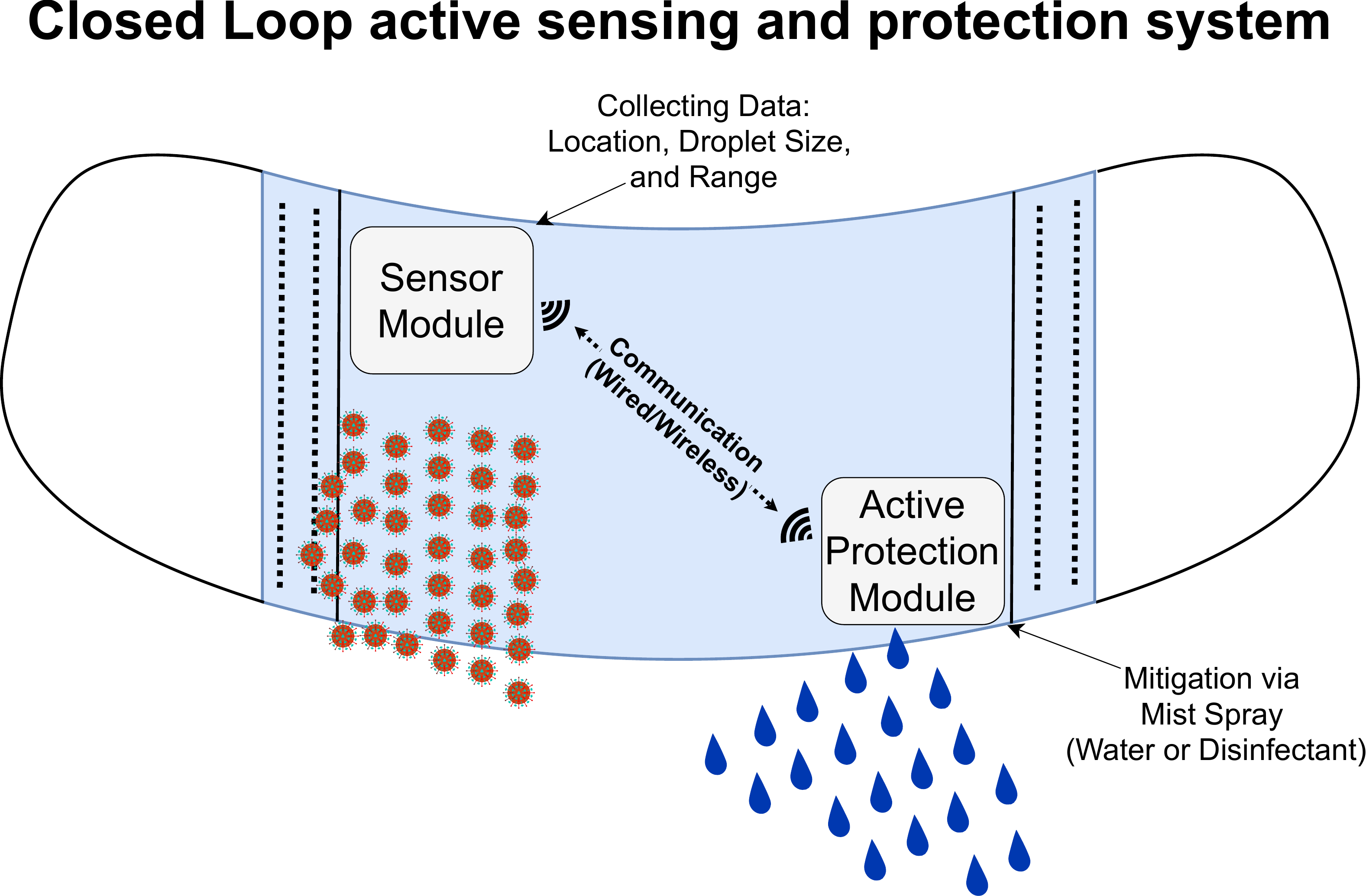}
\caption {Overall architecture of the proposed smart mask.}
\label{fig:overall_block_dia}
\end{figure}

Typical Personal Protective Equipment (PPE), such as face masks (cloth, surgical, or N95), face shields, eye protection, disposable gloves, and coveralls all provide passive protection: these devices only prevent pathogens from entering the body by filtering them out. By contrast, active protection devices can actively attack and destroy pathogens near vulnerable parts of the body (e.g., the nose and mouth). Here we consider closed-loop active or ``smart'' masks for use in places where potentially virus-laden respiratory droplets (typically 0.1-10~$\mu$m in diameter) are most likely to be transmitted; examples include bathrooms, doctor's offices, daycare centers, and public transportation. Fig.~\ref{fig:overall_block_dia} gives an overview of the proposed smart mask architecture. A detailed comparison of existing masks with the proposed smart mask is shown in Table~\ref{tab:comparison}. The table shows that the proposed system detects airborne droplets (potentially containing viruses for COVID-19, influenza, measles, or other diseases) and limits their spread via an appropriate situation-aware mitigation strategy. The chosen strategy should be ``smart'', i.e., adaptive based on sensor outputs that provide information on concentration, size distribution, and other properties of the droplets. In our initial implementation, mitigation is provided by a cold mist generator that loads the droplets (thus making them quickly fall to the ground), and adaptation is provided by algorithms running on an on-board controller that adjust the spray angle, intensity, and duration of the generator based on sensor data. Such active closed-loop protection can remove viruses (and other pathogens) from the air before they infect others, thereby reducing the need for periodic disinfection of the area, while providing increased protection to the wearer.

\begin{table}[]
\centering
\caption{A detailed comparison of key features of various existing masks and the proposed smart mask.}
\label{tab:comparison}
\begin{tabular}{@{}lllll@{}}
\toprule
Properties                                                                       & Cotton mask                                                                                                                          & Surgical mask                                                                                                                                             & N95 Respirator mask                                                                                                                                     & Smart mask                                                                                                                   \\ \midrule
Purpose                                                                          & \begin{tabular}[c]{@{}l@{}}Reduces the amount \\ of expiratory droplets\\ expelled by a person\end{tabular} & \begin{tabular}[c]{@{}l@{}}Wearer gets protection \\ against large droplets \\ and splashes of other’s \\ bodily fluids\end{tabular} & \begin{tabular}[c]{@{}l@{}}Protects wearer by \\ reducing exposure to \\ airborne particles \\ (only non-oil aerosols)\end{tabular} & \begin{tabular}[c]{@{}l@{}}Protects wearer by \\ reducing exposure to\\  airborne particles\end{tabular} \\
Face seal Fit                                                                    & Loose-fitting                                                                                                                        & Loose-fitting                                                                                                                                             & Tight-fitting                                                                                                                                           & Close-fitting                                                                                                                \\
Filtration                                                                       & Low level                                                                                                                            & Moderate level                                                                                                                                            & High level (95\%)                                                                                                                                       & \begin{tabular}[c]{@{}l@{}}Removes all large (dia. $>5$~$\mu$m),  \\ and tiny (dia. $<5$~$\mu$m)  \\ airborne particles/droplets\end{tabular}  \\
Breathability                                                                    & Breathable                                                                                                                           & Breathable                                                                                                                                                & Difficult                                                                                                                                               & Breathable                                                                                                                   \\
\begin{tabular}[c]{@{}l@{}}User seal \\ check requirement\end{tabular}           & No                                                                                                                                   & No                                                                                                                                                        & Yes                                                                                                                                                     & Yes                                                                                                                          \\
Leakage                                                                          & Through cloth                                                                                                                        & Around mask edges                                                                                                                                         & Minimal leakage                                                                                                                                         & Minimal leakage                                                                                                              \\
\begin{tabular}[c]{@{}l@{}}Presence of sensor \\ to detect droplets\end{tabular} & No                                                                                                                                   & No                                                                                                                                                        & No                                                                                                                                                      & Yes                                                                                                                          \\
\begin{tabular}[c]{@{}l@{}}Presence of \\ spraying mechanism\end{tabular}        & No                                                                                                                                   & No                                                                                                                                                        & No                                                                                                                                                      & Yes                                                                                                                          \\
Limitation                                                                       & Wash after each use                                                                                                                  & Discard after each use                                                                                                                                    & \begin{tabular}[c]{@{}l@{}}Ideally discard\\ after each use\end{tabular}                                                                                & \begin{tabular}[c]{@{}l@{}}Reusable and needs to be \\ disinfected after each use\end{tabular}                               \\ \bottomrule
\end{tabular}
\end{table}

The paper is organized as follows. The airborne transmission of viruses is briefly reviewed in Section~\ref{sec:smart_masks}. The system-level architecture and operating principles of smart masks are presented in Sections~\ref{sec:architecture} and \ref{sec:operation}, respectively. The experimental setup, results, and key observations are explained in Sections~\ref{sec:setup}, \ref{sec:results}, and \ref{sec:observations}, respectively. The commercial feasibility of smart masks is discussed in Section~\ref{sec:cost}, and future work and conclusions are presented in Section~\ref{sec:conclusion}.

\begin{figure}[t]
\centering
\includegraphics[width=\linewidth]{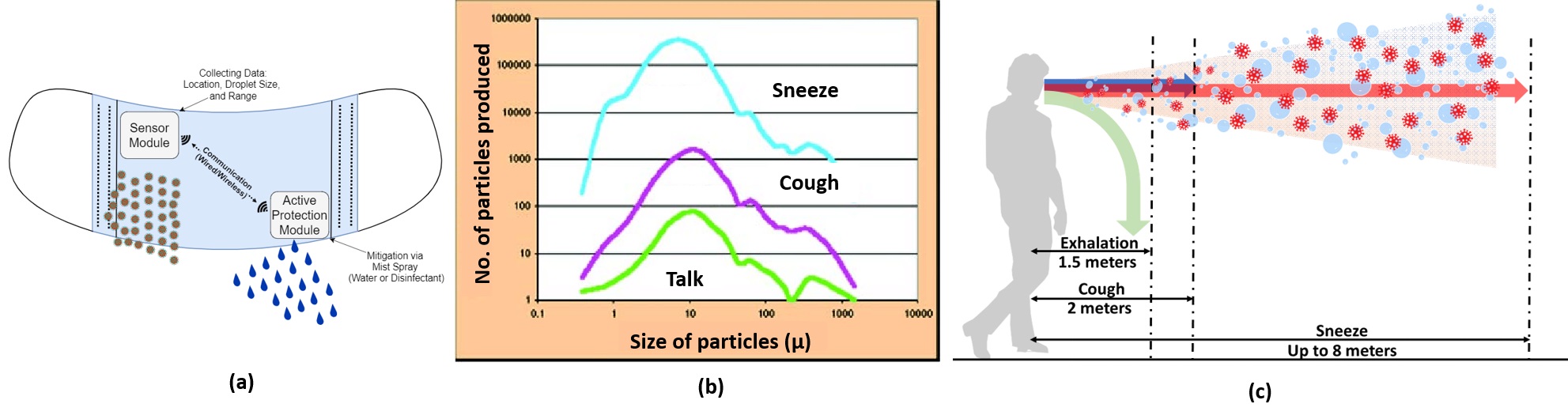}
\caption {(a) Estimated size distribution of particles produced during talking, coughing, and sneezing~\cite{memarzadeh2010improved}. (b) Estimated spread of droplets produced while normal exhalation, coughing, and sneezing. (c) Visual representation of the active protection provided by the proposed smart mask.}
\label{fig:basic_info}
\end{figure}

\section{System Architecture and Implementation}
\label{sec:smart_masks}
Humans naturally generate a variety of droplets (\textit{e.g.}, due to breathing, talking, sneezing, and coughing). These include various cell types (\textit{e.g.}, epithelial and immune system cells), physiological electrolytes contained in mucus and saliva (\textit{e.g.}, Na$^+$, K$^+$, Cl$^-$), and, potentially, various infectious agents (\textit{e.g.}, bacteria, fungi, and viruses). Droplets $>5$~$\mu$m in diameter tend to remain trapped in the upper respiratory tract (the oropharynx, i.e., nose and throat area). By contrast, smaller droplets can be inhaled into the lower respiratory tract (the bronchi and alveoli in the lungs)~\cite{chartier2009natural}.

Infectious aerosols are generated when pathogens from patients' respiratory tracts come into contact with exhaled air and dissolve in any droplets produced by sneezing, coughing, or talking. According to some publications~\cite{memarzadeh2010improved, bourouiba2014violent}, a single sneeze may produce as many as $40,000$ droplets with diameters between 0.5-12~$\mu$m (Fig.~\ref{fig:basic_info}(a)). These droplets may be expelled at speeds up to 100 m/s, and reach distances up to 8 meters (Fig.~\ref{fig:basic_info}(b)). By contrast, a single cough may produce up to $3,000$ droplet nuclei, and similar numbers are typically recorded after talking for five minutes. Air flows can carry infectious droplets long distances; such flows occur even during the most innocuous daily activities, such as walking or opening doors. In addition, differences in temperature (and thus density) across open doorways result in convective currents that exchange air between nearby areas.

Virus particles do not float freely in the air, but are always suspended in droplet nuclei that are significantly larger than the virus itself. The SARS-CoV-2 virus is 100~nm in diameter, and can remain suspended within droplets $>0.2$~$\mu$m in size. Droplets $>5$~$\mu$m fall to the ground quickly, while very small droplets evaporate~\cite{wells1934air} and aerosolize in a few seconds to droplet nuclei $\sim$1~$\mu$m in size~\cite{marr2019mechanistic}. Fortunately, most masks can filter out droplets of this size: many materials have $\geq96\%$ filtration efficacy for particles $>0.3$~$\mu$m, including 600 TPI (threads per inch) cotton, cotton quilts, and cotton layered with chiffon, silk, or flannel~\cite{viola2020face, kumar2020utility}. Thus, the most important requirement for the proposed smart mask  is to eliminate the small (but potentially significant) fraction of virus-laden droplets that are $<0.3$~$\mu$m in size. These small droplets are removed by creating an air-flow pattern close to the mask through spraying a mist. It blows the droplets away from the wearer and also ``loads" them (i.e., increases their mass and size), thus causing them to quickly fall to the ground (Fig.~\ref{fig:basic_info}(c)).

\subsection{System Architecture}
\label{sec:architecture}
The smart mask has two main components: a particulate sensor and an active mitigation device. Methods for sensing airborne pathogens and allergens can be divided into sampling-based (local) and remote detection approaches. A variety of pre-concentration and sampling methods based on solid impactors, liquid impactors, and filters are available~\cite{verreault2008methods}. These methods have the advantage that the sampled pathogens can be analyzed, identified, and quantified using sensitive lab-based techniques, such as real-time polymerase chain reaction (RT-PCR)~\cite{spackman2008type} or surface-enhanced Raman spectroscopy (SERS)~\cite{shanmukh2006rapid, yeh2020rapid}. However, incorporating such sensitive detectors into a wearable form factor is extremely difficult.

Due to this limitation, we focus on a remote detection approach. Remote (also known as stand-off) detection of pathogens has been demonstrated using a variety of optical methods, using asymmetric microsphere resonant cavities~\cite{ballard2015stand}, laser-induced fluorescence~\cite{farsund2010required}, and random Raman lasing~\cite{hokr2014single}, as well as non-optical methods, such as THz imaging and spectroscopy~\cite{galoda2007fighting, keshavarz2019sensing}. Here, we propose a particulate matter (PM) sensor that makes use of laser scattering to precisely count airborne particles in multiple size ``bins''. In addition, the control algorithm also uses data from auxiliary sensors (such as relative humidity and temperature) while determining the optimal parameters for the mitigation device (i.e,. mist generator), since aerosol travel distances depend on such environmental factors~\cite{verreault2008methods}.

The mitigation device generates aerosolized mist on-demand using a piezoelectric transducer. A variety of liquids, including pure water or mixtures of water with various impurities (to increase droplet mass) or disinfectants, can be used to generate the mist. The best disinfectant for a given pathogen can be found using guidelines provided by the U.S. Centers for Disease Control and Prevention (CDC)~\cite{CDC}  and Environmental Protection Agency (EPA)~\cite{United}. Common disinfectants include diluted bleach, soap, and $>70\%$ alcohol solution.

\begin{figure}[t]
\centering
\includegraphics[width=\linewidth]{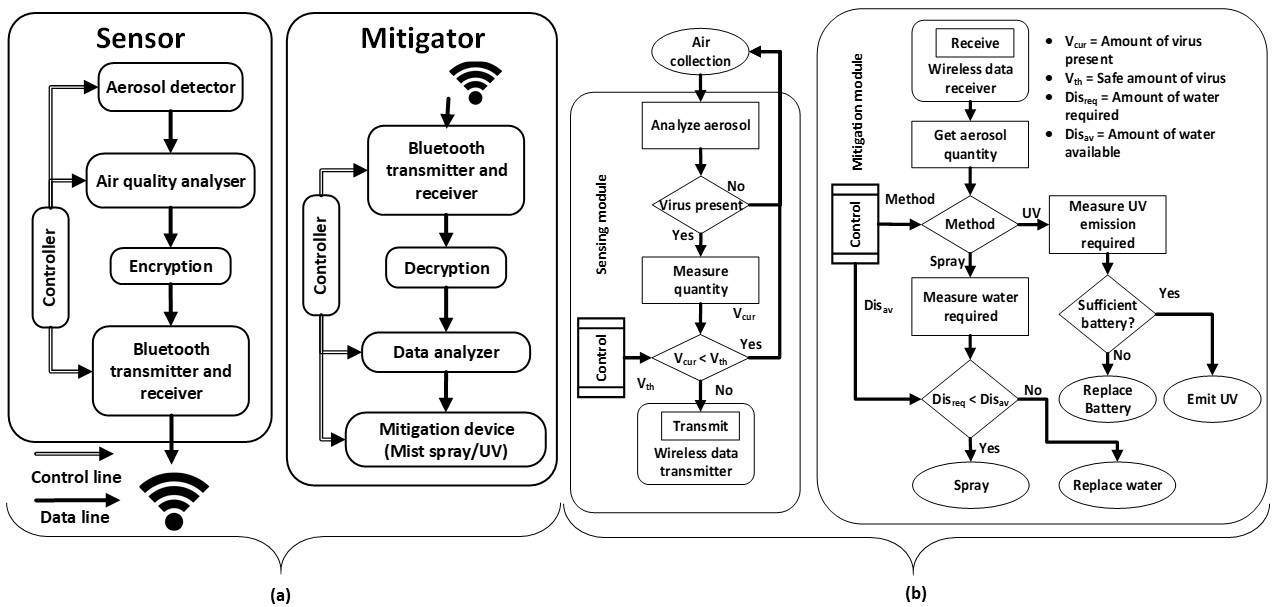}
\caption{(a) System-level block diagram, and (b) operational flowchart of the proposed smart mask.}
\label{fig:block_dia_and_flowchart}
\end{figure}

\subsection{Operational Principles}
\label{sec:operation}
The smart mask integrates a sensing module, as shown in Fig.~\ref{fig:block_dia_and_flowchart}(a), that senses the presence of airborne aerosol droplets (typical diameter 0.1-10~$\mu$m) in the vicinity of the user’s respiratory tract. It incorporates an optical detector system and auxiliary humidity and temperature sensors (as described previously) to quantify the total number and size distribution of these droplets as they approach the protected region (\textit{i.e.}, the nose and mouth). The outputs of both sensors are processed by an air quality analyzer module. The latter uses algorithms to analyze sensor data and thereby classify the quality of the incoming air stream based on health risk (\textit{e.g.}, ``very high'', ``high'', ``moderate'', and ``low''). These risk categories are then encrypted for security and sent to the protection module either wirelessly (\textit{e.g.}, over Bluetooth) or by using a wired connection. A ``high risk'' output triggers the active protection mechanism, as summarized in Fig.~\ref{fig:block_dia_and_flowchart}(b). The protection mechanism sprays a mist that is safe for human exposure by using a micro nozzle. This module is triggered by an electromechanical relay driven by the system controller.

The hardware and software required for both sensing and protection modules are implemented using low-cost commercial off-the-shelf components (a low-power microcontroller and a wireless system-on-module) built within a mask, thus enabling widespread deployment in the vulnerable population. Further, the device can be equipped with machine learning (ML) algorithms that learn when respiratory droplets are likely to be present in a location and proactively employ the proposed active protection mechanism. The smart mask can also connect to authorized mobile devices through its wireless module. Users can use a mobile application to monitor current air quality, check system status (e.g., battery and liquid levels), and also manually override the on-board mitigation algorithm if needed.

\section{Experimental System and Results}
\label{sec:exp_results}

\begin{figure}[t]
\centering
\includegraphics[width=0.9\linewidth]{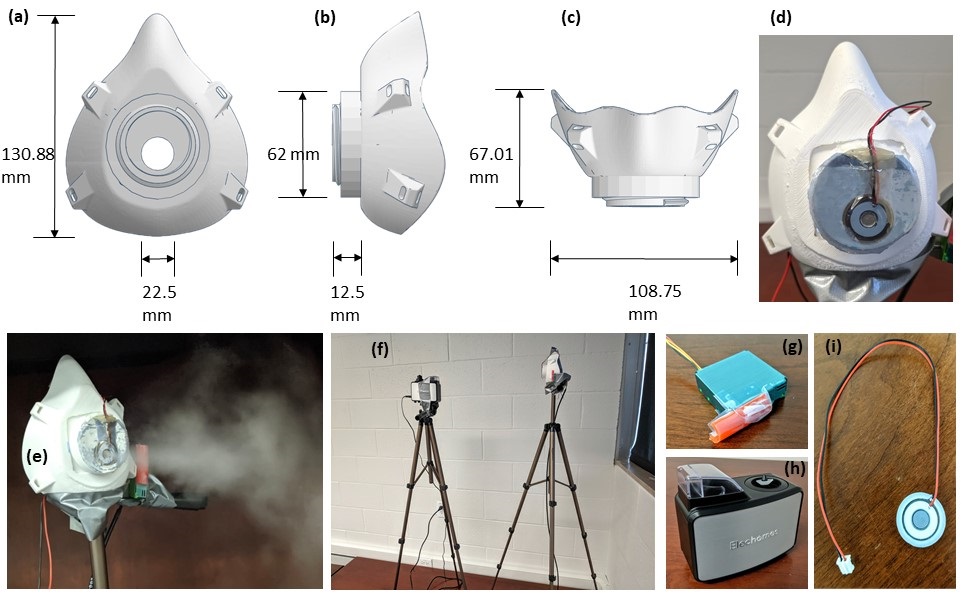}
\caption{(a) Front, (b) side, and (c) top view of the smart mask design, including dimensions. (d) A prototype of the smart mask based on the design in (a)-(c). (e) The smart mask prototype in operation, showing the sensing and mitigation modules. (f) Experimental setup showing the smart mask and humidifier. (g) Particulate matter (PM) sensor used in the experiments. (h) Humidifier used to mimic aerosols produced from human activity. (i) Piezoelectric transducer used in the mist generator.}
\label{fig:setup}
\end{figure}

\subsection{Test Setup}
\label{sec:setup}
3D views of our smart mask prototype are shown in Fig.~\ref{fig:setup}(a)-(c) respectively; a photograph of the fully-assembled version is shown in Fig.~\ref{fig:setup}(d, e). The experimental setup for testing the functional prototype uses a humidifier to replicate aerosol-sized droplets generated during daily activities like talking, coughing, and sneezing. The humidifier (Fig.~\ref{fig:setup}(h)) produces mist droplets in the 0.3-2.5~$\mu$m range. Since standard face masks filter out most large droplets (as mentioned earlier)~\cite{viola2020face, kumar2020utility}, we check the smart mask's performance against droplets of size $<0.3$~$\mu$m. We use two PM sensors (Sensirion SPS30, see Fig.~\ref{fig:setup}(g)): one on the smart mask (which is placed at a height of $\sim$1.6~m), and the other on the ground (Fig.~\ref{fig:setup}(f)).  The use of two sensors allows us to quantify both predicted effects of PM loading by the mitigation spray: i) reduction in PM count near the mask, and ii) increase in PM count on the ground (due to the rapid falling of loaded particles). Note that the PM sensor on the ground is not part of the mask, and only used during testing. The chosen PM sensor is based on laser scattering and has built-in ``contamination resistance'' feature in which artificial air flow (created using a small built-in fan) is used to drive deposited particles out of the sensor. Due to the placement of the exhaust vent, this air flow interferes with the incoming air. To reduce this effect, we provided a guided path for the exhaust as shown in Fig.~\ref{fig:setup}(g).

The smart mitigation module uses a piezoelectric transducer (Fig.~\ref{fig:setup}(i)) that vibrates at a frequency of 110~kHz. The vibrating portion of the transducer is a mesh-like structure with one side facing the liquid and the other side facing the atmosphere. The pressure drop created by the vibration converts liquid water into vapor, which exits the transducer as shown in Fig.~\ref{fig:setup}(d). The transducer runs at 12~V while consuming up to 0.2~A of current. Power is provided by four 18650 Li-ion cells connected in series with a total capacity of 2000~mAh, which is sufficient for a typical day of heavy use. All experiments were conducted in a well-controlled indoor environment with no nearby air vents; this is because aerosols are very sensitive to weak air flows (e.g., due to air-conditioning).

\begin{figure}[t]
\centering
\includegraphics[width=\linewidth]{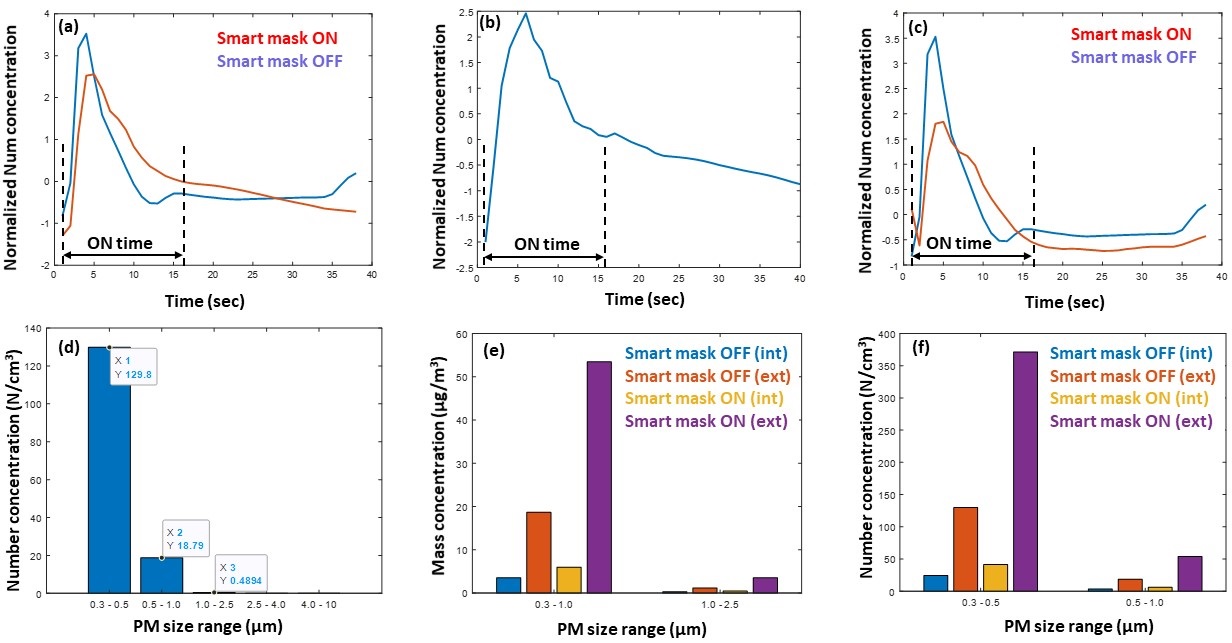}
\caption{The PM count for sizes 0.3-1.0~$\mu$m remaining near the smart mask: (a) decreases when the mitigation module is used, (b) is increased by the aerosol droplets generated by the smart mask itself, and (c) decreases even more significantly when the self-generated PM count is subtracted. (d) The PM number concentration produced by the humidifier when set to replicate daily activities like talking, coughing, and sneezing. Comparison of (e) mass, and (b) number concentrations from the two PM sensors (int: on the mask, and ext: on the ground) when the smart mask's mitigation function is turned ON and OFF.}
\label{fig:results}
\end{figure}

\subsection{Experimental Results}
\label{sec:results}
Initially, the outputs of the two sensors were checked to ensure a stable and uniform PM concentration in the area around the experimental setup. Next, we performed two calibration experiments. First, the humidifier was turned ON, which imitates droplets produced by human actions, for $15$~sec with the mitigation module turned OFF. The aerosols were then allowed to settle for $160$~sec, with both PM sensors recording their local number and mass concentrations. Second, the same experiment was repeated, but with the smart mask's mitigation module turned ON under normal conditions \textit{i.e.}, with the humidifier turned OFF. Finally, the effectiveness of the smart mask was verified, as follows. The humidifier was turned ON for $15$~sec, and the smart mask activated for 15~sec once its sensor detects a significant local change in PM concentration. The outputs of both PM sensors were then monitored until all aerosols settle out ($\sim$160~sec). The results of these experiments are analyzed in the next section.

\subsection{Key Observations}
\label{sec:observations}
Fig.~\ref{fig:results}(a) shows PM number concentration detected by the sensor on the smart mask, with only the size range of interest (0.3-1.0~$\mu$m) considered, in two cases: mitigation module ON and OFF. Clearly, the mitigation module (mist generator) significantly reduces the local PM number within this critical size range -- simple analysis suggests a time-averaged reduction of $\sim$20\%. Moreover, the PM count with the mitigation module ON is artificially increased due to the aerosols generated by the module itself (Fig.~\ref{fig:results}(b)). Once this self-interference term is subtracted, the time-averaged reduction in PM count due to the mask becomes even more significant -- Fig.~\ref{fig:results}(c) suggests $\sim$40\%.

The aerosol concentrations produced by the humidifier in the 0.3-2.5~$\mu$m range resemble those produced during daily activities, and are shown in Fig.~\ref{fig:results}(d). Next, we analyze the effect of the smart mask on the concentration of these aerosolized particles. Figs.~\ref{fig:results}(e) and (f) compare the time-averaged mass and number concentrations with the mask ON and OFF for both internal (on the mask) and external (on the ground) PM sensors. The data shows that the mass concentrations measured by the external sensor in the 0.3-1.0~$\mu$m and 1.0-2.5~$\mu$m ranges increase by $\sim$63\% and $\sim$60\%, respectively, when the mask is turned ON. Similarly, the corresponding number concentrations increase by $\sim$62\% and $\sim$50\%, respectively. These results suggest that the smart mask is loading the local aerosol particles, as expected, thus causing them to fall faster to the ground.

\subsection{Mechanical Design and Cost Analysis}
\label{sec:cost}
The mechanical structure of the mask is based on a open-source 3D design, but was modified to include a compartment for liquid storage. This compartment has two holes: one for refilling and the other for the transducer. The transducer is placed flush with the mask's outer surface, with one of its sides facing the liquid. A plastic layer is used to ensure a watertight seal around the transducer. Small perforations backed by replaceable air filters are added on the sides of the mask to ensure adequate air intake. Additionally, the edges are lined with silicone, thus ensuring a good seal with the skin. A second compartment on the mask holds the sensor and wiring harness, while a small pocket- or belt-worn external unit  houses the battery and system controller. The mask was 3D-printed for testing purposes, but can instead be injection-molded to reduce cost for bulk production. In this case, the overall manufacturing cost per mask (including the electronics) is expected to be near \$25 based on the bill of materials.

\subsection{Integration with smartphone}
\label{sec:integration}
The integration of the mask with a smartphone is a vital part of the project. It helps in the transfer and visualization of the data, changing modes, and controlling the smart mask. We developed an android application that can connect to multiple smart masks at the same time. It collects data from the smart mask using Bluetooth connection. The mask sends important particulate matter data to the application which is processed and shown to the user in the form of real-time plots. Using the transferred data, the application can also suggest the user decontaminate the mask in the form of a notification when the number of particles the mask is exposed to exceeds a threshold. The application can control the switching of the mask between turn-ON and turn-OFF. Also, the application helps in choosing different functioning modes of the mask: automatic, and manual. In manual mode, the user has the power of when to turn-ON or turn-OFF the mask. In the automatic mode, the smart mask automatically switches based on the particulate matter density in its surroundings.

\section{Conclusion and Future Work}
\label{sec:conclusion}
We have presented a new paradigm of active closed-loop defense, in the form of a smart mask, against airborne pathogens including SARS-CoV-2. We have presented the system design, and using a functional prototype of the mask, also demonstrated the major operational characteristics. Various levels of smartness can be incorporated into the proposed system to control the time, duration, and intensity of mitigation based on awareness of the location (e.g., hospitals, quarantined zones, or care centers with infected patients), ambient conditions (e.g., humidity and air temperature, human occupancy), and overall health of the user (e.g., age, pre-existing conditions, etc.). The proposed mask can also be extended to protect against pollutants, dust particles, and pollen, e.g., for vulnerable populations with pollen and/or dust allergies. It can also be extended to other usage scenarios, e.g., military personnel exposed to harmful airborne particles; dentists performing dental procedures; and day-care or elementary schools where social distancing is hard to maintain. While more testing and evaluation is needed to fully establish the merits of the smart mask and identify remaining design challenges and associated trade-offs, the initial results are highly promising. The proposed defense can be applied to existing masks as an add-on reusable assembly, as well as to new mask designs as demonstrated here. It also has the potential to completely replace traditional masks for specific applications. Our future studies will focus on these topics.

\bibliographystyle{IEEEtran}
\bibliography{main}

\end{document}